%% file: scaling.tex
\newif\ifdraft
\newif\iffull
\newif\ifcomment
\drafttrue
\fulltrue
\def\dvers{v0.1.1}
\documentclass[review]{elsarticle}

\usepackage{lineno,hyperref}
\modulolinenumbers[5]

\usepackage{graphicx}  
\usepackage{dcolumn}   
\usepackage{bm}        
\usepackage{amssymb}   
\usepackage{amsfonts}
\usepackage{graphics}
\usepackage{epsfig}
\usepackage{units}
\usepackage[usenames]{color}

\graphicspath{{./img/}}
\ifdraft
\usepackage{lineno}
\linenumbers
\modulolinenumbers[5]
\usepackage{fancyhdr}
\else

\fi

\begin{document}
\title{Geometrical Scaling of Direct-Photon Production in Hadron Collisions from RHIC to the LHC}
\iffull
\author[wwu,emmi]{Christian Klein-Boesing}
\ead{Christian.Klein-Boesing@wwu.de}
\author[bnl,rbrc,ccnu]{Larry McLerran}
\ead{mclerran@bnl.gov}
\address[wwu]{Institut f{\"u}r Kernphysik -  M{\"u}nster, Germany}
\address[emmi]{ExtreMe Matter Institute, GSI - Darmstadt, Germany}
\address[bnl]{Physics Dept, Bdg. 510A, Brookhaven National Laboratory, Upton, NY-11973, USA}
\address[rbrc]{RIKEN BNL Research Center, Brookhaven National Laboratory, Upton NY 11973, USA}
\address[ccnu]{Physics Dept, China Central Normal University, Wuhan, China}
\else
\fi
\vspace{0.3cm}
\ifdraft
\date{\today, \color{red}DRAFT \dvers\ \color{black}}
\else
\date{\today}
\fi
\begin{abstract}
Geometric scaling is a property of hadronic interactions 
predicted by theories of gluon saturation and expresses rates in terms of dimensionless ratios
of transverse momentum to the saturation momentum. In this paper we consider production of photons in $pp$, $dAu$ and $AuAu$ collisions at $\sqrt{s_{NN}} = 200$\,GeV (RHIC) and in $PbPb$ collisions at
$\sqrt{s_{NN}} = 2760$\,GeV (LHC) and show that the yield of direct photons in the transverse momentum range
$1$~GeV$ < p_T \le 4$~GeV/$c$ satisfies geometric scaling.   Excellent agreement with geometric scaling
is obtained with the only free parameter of the saturation momentum
determined previously via the dependence of the saturation momentum upon Bjorken $x$ and centrality.
\end{abstract}
\maketitle
\ifdraft
\thispagestyle{fancyplain}
\fi
\input{scalingMain}

\iffull
\fi
\bibliographystyle{apsrev4-1}
\bibliography{master,servant}
\end{document}

%% file: scalingMain.tex


\linenumbers

\section{Introduction}

The phenomenon of gluon saturation arises at high energies when the density of gluons per unit area in a hadron is large \cite{Gribov:1984tu,Mueller:1985wy,McLerran:1993ni,McLerran:1993ka}.  It implies the existence of a saturation momentum scale:
\begin{equation}
  Q_{sat}^2 = {\kappa \over {\pi R^2}} {{dN} \over {dy}},
\end{equation}
where $R$ is the hadron size, $dN/dy$ is the gluon density per unit rapidity, and $\kappa$ is a constant of order 1.  Up to effects of a running coupling constant, at very large $Q_{sat}$,
the saturation momentum is the only scale for physical processes.   This implies scaling relations for physical processes.  In particular, geometric scaling 
was first discovered in deep-inelastic scattering \cite{Stasto:2000er,Iancu:2002tr}.  It was later applied to high energy particle production in hadron-hadron scattering and explains features of $pp$ and $pA$ scattering as a function of multiplicity, as well as particle production in heavy-ion collisions for fixed centrality as a function of energy \cite{McLerran:2010ex,Praszalowicz:2011tc,Praszalowicz:2011rm,Praszalowicz:2012ab,McLerran:2013oju}.

In this paper, we intend to apply geometric scaling to photon production in hadron-hadron scattering at RHIC and LHC energies ($\sqrt{s_{NN}} = 200$~GeV and 2.76 TeV).  This is an extension of work
where geometric scaling was applied to reproduce the multiplicity dependence of photon production data for AuAu collisions at RHIC \cite{Chiu:2012ij}.  This paper considers in addition $pp$ and $dAu$ collisions at RHIC energy and $PbPb$ collisions at LHC energy.  The obtained agreement with experimental data indicates that geometric scaling works well for photon production.


\section{Scaling}

Geometric scaling is a property of particle densities.  In the theory
of the Color Glass Condensate, one computes these densities from an
underlying theory. In the absence of the effect of running coupling, this theory is controlled by only one scale, the saturation momentum.  Therefore, in a collision with overlap
area $\pi R^2$, for the production of a particle (photon) of momentum $p_T$:
\begin{equation}
\label{eq:yield}
	\frac{1}{\pi R^2} \frac{d^2N}{dyd^2p_T} = F \left(\frac{Q_{sat}}{p_T}\right).
\end{equation}
The transverse overlap area $\pi R^2$ can be estimated for symmetric systems to be proportional to $N_{part}^{2/3}$ \cite{Kharzeev:2000ph}. The saturation scale is given by \cite{Praszalowicz:2011rm,Kharzeev:2000ph}:
\begin{equation}
\label{eq:qsat}
  Q_{sat}^2 = Q_0^2 \cdot N_{part}^{1/3} \left(\frac{E}{p_T}\right)^\delta
\end{equation}
with $\delta$ in the range of 0.22 to 0.28, $Q_0$ of the order of 1\,GeV, and $E$ the center of mass energy $\sqrt{s_{NN}}$.    This parameterization is consistent with fits to deep inelastic scattering \cite{Kowalski:2007rw}.

\begin{figure}[t]
\unitlength\textwidth
\includegraphics[width=1.0\linewidth]{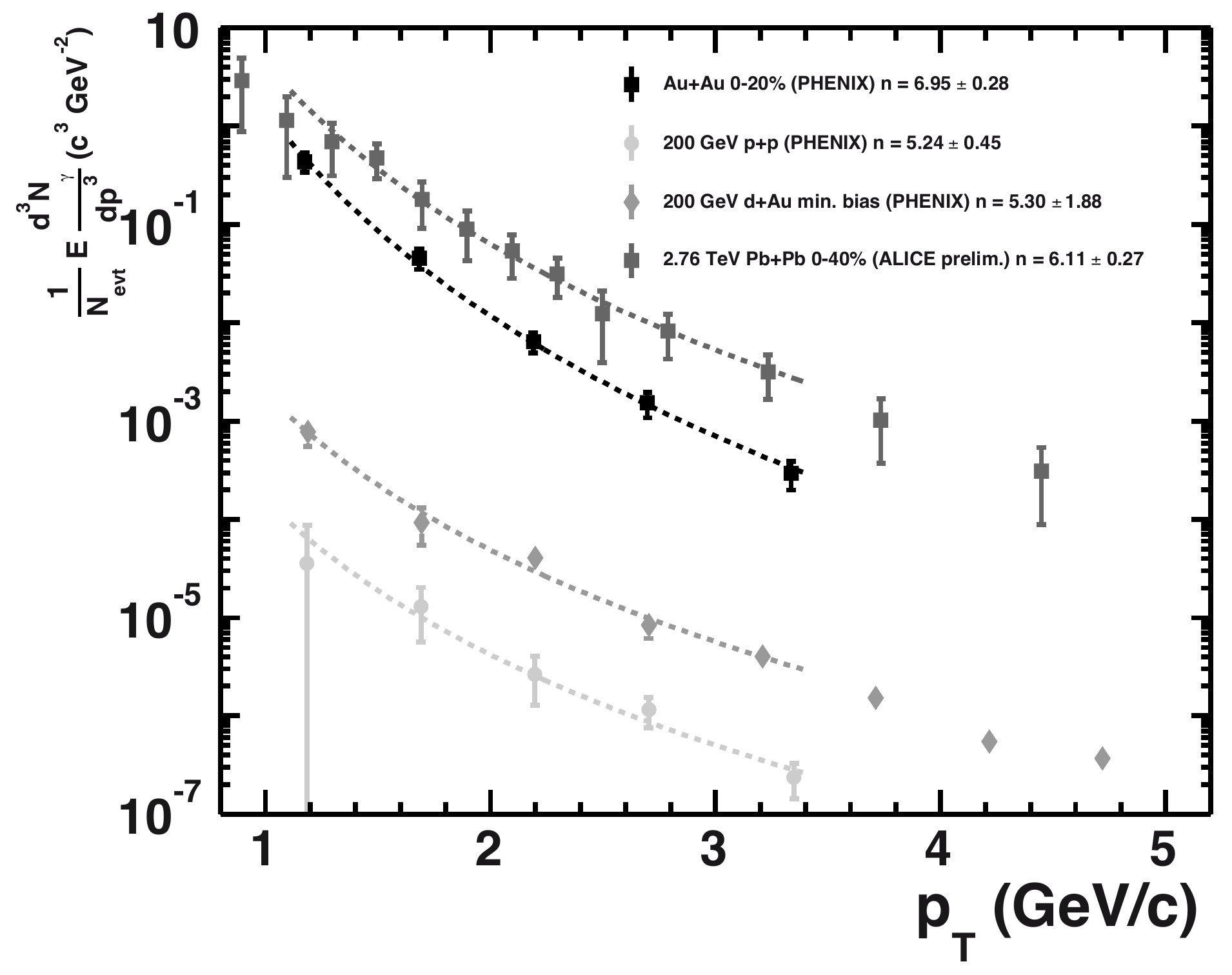}
\caption{Measurements of invariant yields of direct-photon production in nuclear collisions below $p_T = 5$~GeV/$c$ compared to power law parameterizations. Data are taken from the PHENIX experiment at RHIC \cite{Adare:2008ab,Adare:2012vn} and the ALICE experiment at the LHC \cite{Wilde:2012wc}. The error bars represent the combined systematic and statistical uncertainties of the measurements.
\label{fig:spectra} 
}
\end{figure}

The scaling relationship above will work for any function.  It is convenient for us however to parameterize the functional form of the photon distribution
as a power law in $p_T$.  For the finite range of momenta involved, roughly $1-4$~GeV/$c$, such a parameterization of the data is quite good.
We use
\begin{equation}
\label{eq:powerlaw}
 F\propto \left(\frac{Q_{sat}}{{p_T}}\right)^a = \left(\frac{N_{part}^{1/6}\cdot E^{\delta/2}}{p_T^{1+\delta/2}}\right)^a.
 \end{equation}
The geometric scaling assumption can then be tested via rescaling the
invariant yield only, as derived below. Figure~\ref{fig:spectra} shows
a collection of the invariant yields of direct photons measured in
nuclear collisions at $\sqrt{s_{NN}} = 200$ GeV
\cite{Adare:2008ab,Adare:2012vn} and 2.76\,TeV \cite{Wilde:2012wc}. All data have been fit to a power law $A \cdot p_T^{-n}$ and different slopes are extracted for the various systems, between $\approx 5.2 - 6.9$. In the following a slope of $n = 6.1$ will be used.

The knowledge of the slope fixes the only unknown $a$,  when combining Eqs.~(\ref{eq:qsat}) and (\ref{eq:powerlaw}) to extract the $N_{part}$ dependence of the spectrum at a fixed $p_T$: 
\begin{eqnarray}
\label{eq:npart1}
\frac{d^2N}{dyd^2p_T} & = & A p_T^{-n} \propto \left(\frac{N_{part}^{1/6}\cdot E^{\delta/2}}{p_T^{1+\delta/2}}\right)^a \cdot N_{part}^{2/3}\\
\Rightarrow   a & = & n/(1 + \delta/2).
\end{eqnarray}
Taking the full range of $\delta$ and $n$ values as stated above, the invariant yield roughly changes as: 
\begin{eqnarray}
\label{eq:npart2}
\frac{d^2N}{dyd^2p_T} \propto  N_{part}^{1.43 - 1.70}.
 \end{eqnarray}
Using our default values for the slope $n  = 6.1$ and $\delta = 0.25$ it is  $N_{part}^{1.57}$. This estimate is close to the measured  centrality 
dependence of integrated direct-photon yields in $AuAu$ collisions at RHIC, which vary with $N_{part}$ with the  power  of ${1.48\pm 0.08 \mathrm{(stat)} \pm 0.04 \mathrm{(syst)}}$ \cite{Adare:2014aqa}. 

\begin{figure}
\unitlength\textwidth
\includegraphics[width=1.0\linewidth]{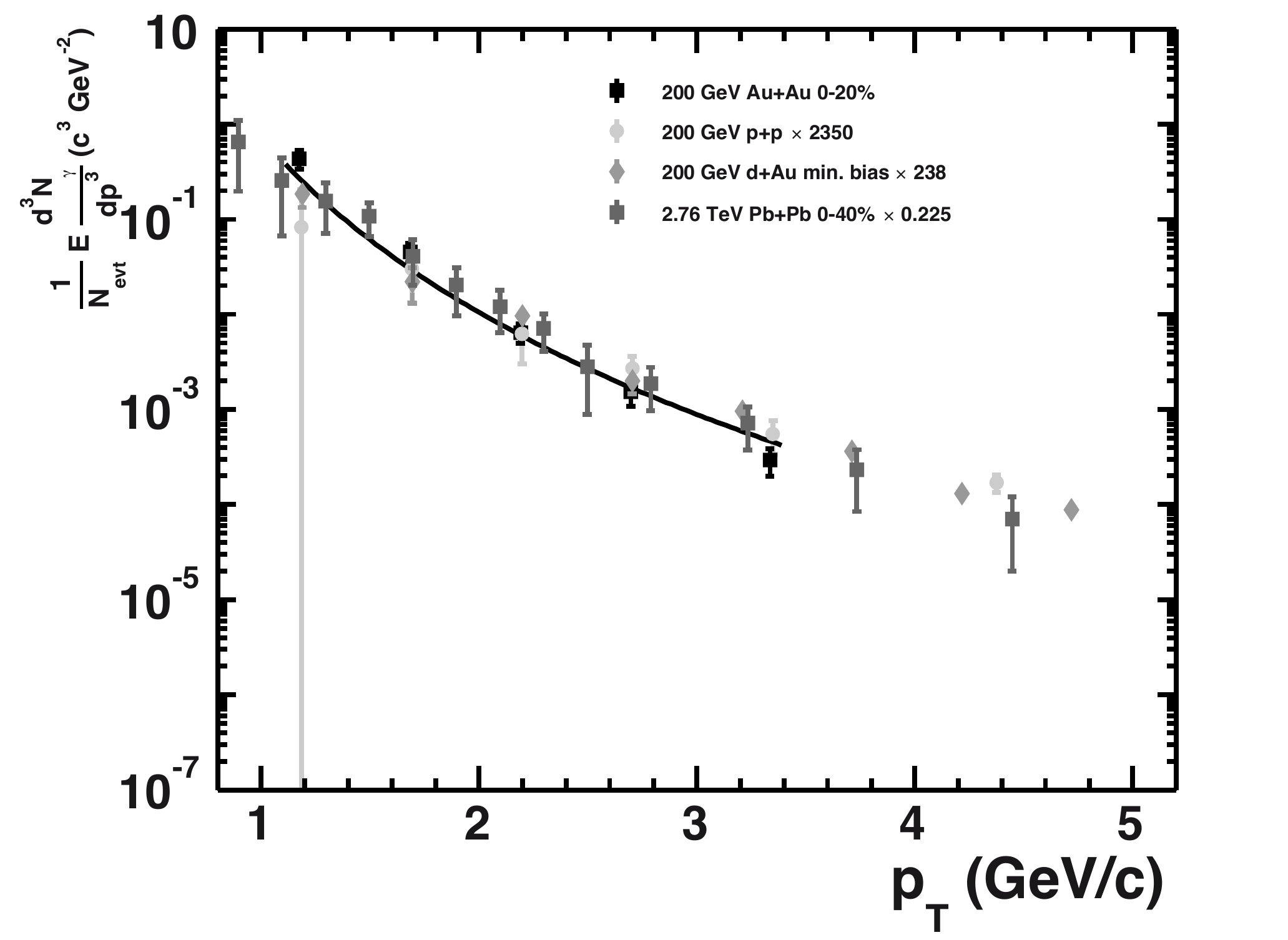}
\caption{Geometrically scaled invariant yields of direct-photon
  production below $p_T = 5$~GeV/$c$ in nuclear collisions at
  $\sqrt{s_{NN}} = 200$ GeV and above. The assumed common power law
  shape of $p_T^{-6.1}$ has been fit to the PHENIX $AuAu$ data and is indicated as black line. The error bars represent the combined systematic and statistical uncertainties of the measurements.  
\label{fig:spectraScaled_200} 
}
\end{figure}

\begin{figure}
\unitlength\textwidth
\includegraphics[width=1.0\linewidth]{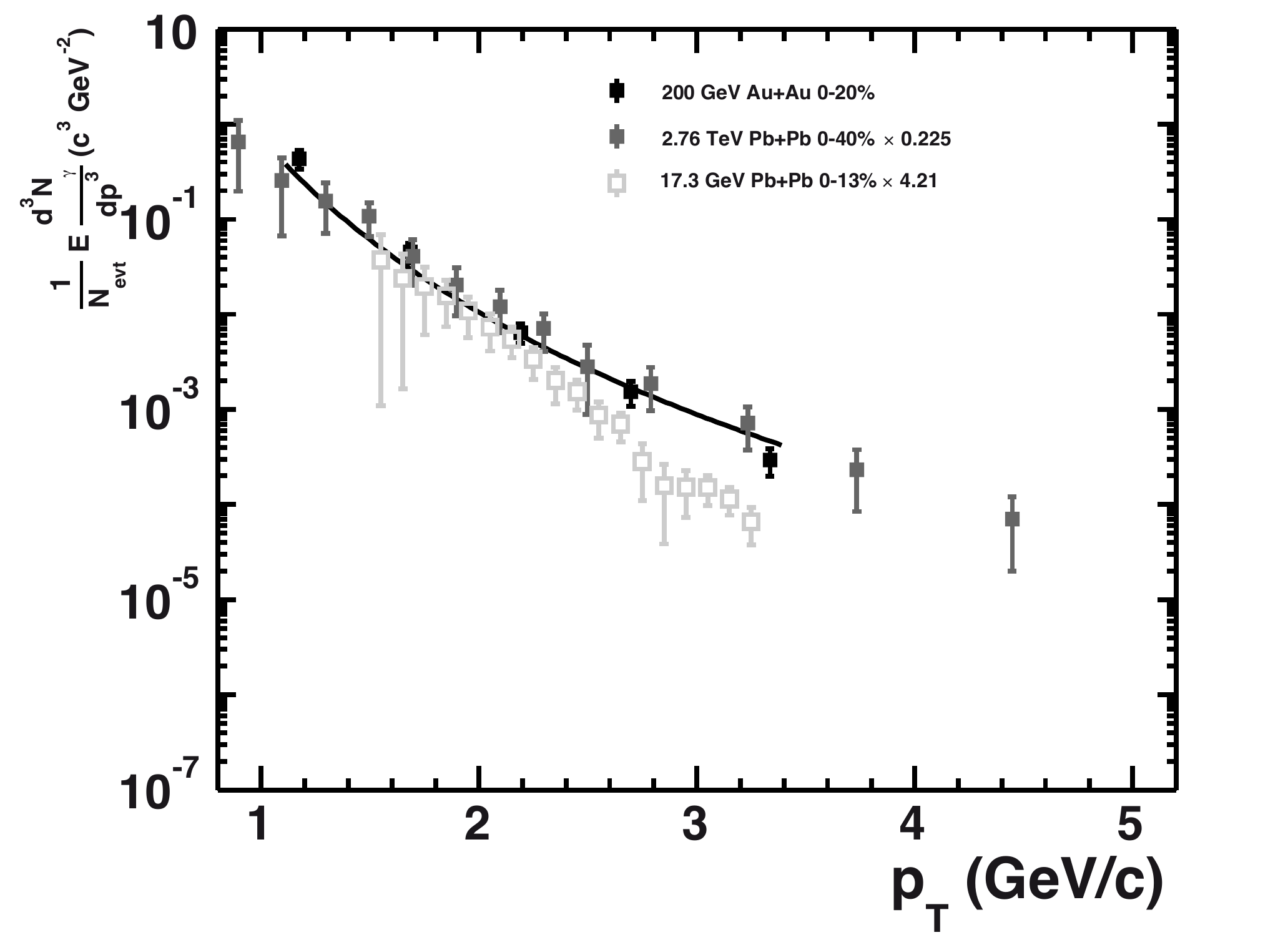}
\caption{Geometrically scaled invariant yields of direct-photon
  production below $p_T = 5$~GeV/$c$ in central heavy-ion collisions
  from $\sqrt{s_{NN}} = 17.3$ GeV to 2.76 TeV. The data are scaled to
  200 GeV following Eq.~(\ref{eq:scaling_aa}). The assumed common
  power law shape is the same as in
  Figure~\ref{fig:spectraScaled_200}. The error bars represent the
  combined systematic and statistical uncertainties of the
  measurements. At SPS energies the geometric scaling is not expected
  to hold, indeed in central $PbPb$ collisions at $\sqrt{s_{NN}} = 17.3$ \cite{Aggarwal:2000ps,Aggarwal:2000th} it is clearly broken above $p_T = 2$ GeV/$c$.  
\label{fig:spectraScaled_hic} 
}
\end{figure}

\begin{figure}[t]
\unitlength\textwidth
\includegraphics[width=1.0\linewidth]{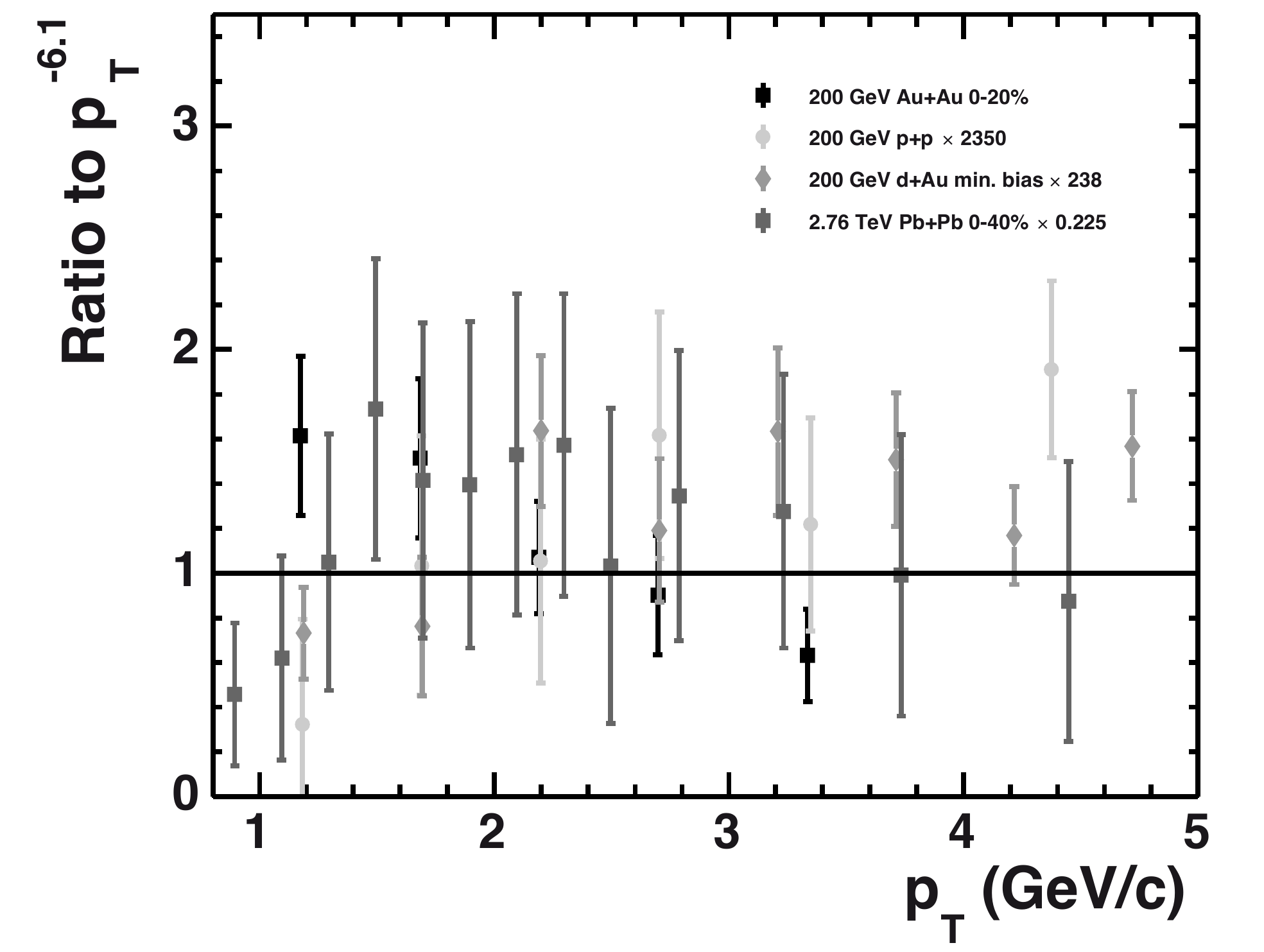}
\caption{Ratio of scaled invariant yields of direct-photon production and the power law used for the scaling (fixed slope of 6.1, fit to the the PHENIX $AuAu$ data). The error bars represent the combined systematic and statistical uncertainties of the measurements.
\label{fig:ratioScaled} 
}
\end{figure}

A general scaling relation between different centralities and/or collision energies is given by the factor
\begin{eqnarray}
\label{eq:scaling_aa}
 \frac{
N_{part,A}^{a/6 + 2/3} \cdot E_{A}^{a\delta/2}
}
{
{\overline N}_{part,A}^{a/6 + 2/3} \cdot {\overline E}_{A}^{a\delta/2}. 
}
\end{eqnarray}
This relation holds for symmetric systems and has been used in
Figure~\ref{fig:spectraScaled_200} to rescale the direct-photon
production in central $PbPb$ collisions at the LHC, as well as in $pp$
at RHIC to the direct-photon production in central $AuAu$ collisions
at RHIC.  In particular it is remarkable that  the measurement of
direct photons in $pp$ obeys the scaling over three orders of
magnitude within less than a factor of two, as seen on the linear
scale in Figure~\ref{fig:ratioScaled}. 
At SPS energies geometric scaling is not expected to hold. For $p_T < 2$\,GeV/$c$ 
we found that the scaled direct-photon data  from central $PbPb$
collisions at $\sqrt{s_{NN}} = 17.3$~GeV \cite{Aggarwal:2000ps,Aggarwal:2000th} is close to the universal curve, while for higher $p_T$ the scaling is clearly broken. The comparison of all direct-photon measurements in heavy-ion collisions, rescaled to central $AuAu$ collisions at RHIC,  is shown in Figure~\ref{fig:spectraScaled_hic}.

For asymmetric systems, such as $dAu$, the scaling  relation is more
complicated, since one cannot use $N_{part}$ any more as a proxy for
the geometry. E.g. in Eq.~(\ref{eq:yield}) only the overlap is
relevant, while $N_{part}$ is largely driven by the thickness of the
larger partner. Due to the asymmetric nature of the deuteron in itself
this overlap area can range between one to two times the $pp$
value. In the following the average number of participants from the
deuteron $\langle N_{part}[d]\rangle = 1.62 \pm 0.01$ as calculated in
\cite{Adare:2013nff} is used as an estimate, so the total overlap area
$\pi R^2$ is
proportional to $3.2^{2/3}$.
Similarly, in Eq.~(\ref{eq:qsat}) the saturation scales of the individual partners need to be considered for $dAu$. The scaling factor between symmetric $AuAu$ collisions and asymmetric, minimum bias $dAu$ collisions is thus estimated as:
\begin{eqnarray}
\label{eq:dau}
\left.\frac{d^2N}{dyd^2p_T}\right|_{AuAu} = \left.\frac{d^2N}{dyd^2p_T}\right|_{dAu} \cdot
\frac{
N_{part,AuAu}^{a/6 + 2/3}
}
{
3.2^{2/3} \cdot 1.6^{a/12} \cdot 197^{a/12}
}.
\end{eqnarray}
Here, the first term of the denominator parameterizes the overlap area in $dAu$ reactions, while for the latter two terms we follow the discussion in \cite{Dumitru:2001ux} and assume that the saturation momentum for the asymmetric $dA$ collision is: 
\begin{equation}
Q_{sat}^2 = \sqrt{Q_{sat,d}^2 Q_{sat,A}^2}.
\end{equation} 
This is the case for an emission energy of the photon large compared to the saturation momentum, which should be the case for $dAu$ collisions at RHIC energies. As discussed in \cite{Kharzeev:2000ph} the saturation scale for nuclei changes with the length scale $A^{1/3}$, however in the case of asymmetric nuclei this length scale is reduced, since only the size of the nucleon along the boost direction is relevant. We have chosen $\langle N_{part}[d]\rangle^{1/3}$ as the effective length of the deuteron. 

The direct-photon yield measured in minimum bias $dAu$ has
been rescaled according to Eq.~(\ref{eq:dau}). Again, we find a
remarkable agreement with the direct photon yield in
central $AuAu$ reactions at the same energy after geometric scaling,
despite a scaling over two orders of magnitude and a very different
scaling law for asymmetric systems (see Figures~\ref{fig:spectraScaled_200} and \ref{fig:ratioScaled}).

 \section{Summary and Conclusions}

We have shown that geometric scaling provides a good description of the energy dependence of photon production in nucleus-nucleus collisions, including $pp$ and $dAu$ scattering (Table~\ref{tab:part}).  This is quite remarkable since this involves an extrapolation over several orders of magnitude in the number of nucleon participants, and because the scaling law for the saturation momentum in $dA$ collisions is different in terms of the number of nucleon participants than it is in symmetric collisions.  

But how can geometric scaling work so well?  It is a property of particle emission that ignores final state interactions, but in particular in heavy-ion collisions one expects that the photons arise from quarks and gluons that have undergone interactions (thermalized).  On the other hand, if there is scale invariance of the expansion, the saturation momentum will remain the only scale in the problem.  In hydrodynamic expansion, this is true in the early stages of the reaction.  At some time however, the expansion of the system in the transverse direction becomes important and there is another
scale in the problem, the size of the nucleus.  At even later times, the system has cooled enough so that hadronic mass scales are important
for decay processes, and again these processes should violate the scaling. 

Thus our observation indicates that direct-photon production occurs mainly before the scale breaking effects of particle masses and system-size become important.  The former would be true if the
system produces photons at an energy scale large compared to meson masses, which might be possible. The latter is more difficult,
since flow measurements for photons demonstrate \cite{Adare:2011zr,Lohner:2012ct} that they do have an azimuthal anisotropy with respect to the event reaction plane. This is conventionally associated with transverse expansion and it
requires that the photons be produced at times where the size of the system actually is important.  

So there is a mystery:  How do we maintain geometric scaling in the presence of transverse flow?  If this is possible, it may only be established
after detailed computation that includes the effects of transverse flow.  It also would probably require that the internal dynamics, if associated with early time emission of the glasma would be different from that of the thermalized quark-gluon plasma.  This may be possible, but again requires explicit computation.

Nevertheless, geometric scaling appears to provide an excellent description of the data.  Either it implies there is something very interesting and not yet understood about the dynamics and evolution of the glasma or thermalized quark-gluon plasma, or it is an accident. This result certainly encourages further attempts for a deeper understanding of photon production in this kinematic regime.

\section*{Acknowledgements}
 
 The research of L. McLerran is supported under DOE Contract No. DE-
AC02-98CH10886. The research of C. Klein-B{\"o}sing is supported by the Alliance Program of the
Helmholtz Association (HA216/EMMI). The authors thank Johanna Stachel and Klaus Reygers for the organization of the EMMI Rapid Reaction Task Force \emph{Direct-Photon Flow Puzzle} and the invitation to GSI, where this work has been started.


\begin{table}
\caption{\label{tab:npart}
Employed $N_{part}$ values with the references to the experimental papers on the direct-photon spectra and numerical values on $N_{part}$.}
\begin{tabular}{ccccc}
\hline\hline
$\sqrt{s_{NN}}$ (GeV) &  System & $N_{part}$ & Experiment & References\\
\hline
200 & $p+p$ &  2 & PHENIX &    \cite{Adare:2012vn} \\
200 & $d+$Au & $N^d_{part} = 1.6$ & PHENIX & \cite{Adare:2012vn,Adare:2013nff}\\
200 & Au$+$Au (0-20\%) & 280 & PHENIX & \cite{Adare:2008ab} \\ 
2760 & Pb$+$Pb (0-40\%) & 233 & ALICE & \cite{Wilde:2012wc,Abelev:2013qoq} \\
17.3 & Pb$+$Pb (0-13\%) & 322.5 & WA98 & \cite{Aggarwal:2000ps,Aggarwal:2000th,Aggarwal:2007gw} \\
\hline\hline
\end{tabular} 
\end{table}

